\documentclass[12pt,nofootinbib]{revtex4}

\usepackage{graphicx}
\usepackage{bm}

\newcommand{\beq}{\begin{equation}}
\newcommand{\beqa}{\begin{eqnarray}}
\newcommand{\eeq}{\end{equation}}
\newcommand{\eeqa}{\end{eqnarray}}
\newcommand{\simg}{\gtrsim}
\newcommand{\siml}{\lesssim}

\begin{document}
\title{Reheating After Quintessential Inflation and Gravitational Waves}

\author{Hiroyuki Tashiro\footnote{E-mail: htashiro@tap.scphys.kyoto-u.ac.jp. }
and Takeshi Chiba\footnote{E-mail: chiba@tap.scphys.kyoto-u.ac.jp}}

\address{
Department of Physics, Kyoto University,
Kyoto 606-8502, Japan}

\author{Misao Sasaki\footnote{E-mail: misao@yukawa.kyoto-u.ac.jp}}
\address{
Yukawa Institute for Theoretical Physics, 
Kyoto University,
Kyoto 606-8502, Japan}

\begin{abstract}
We investigate the dependence of the gravitational wave spectrum
from quintessential inflation on the reheating process. 
We consider two extreme reheating processes. 
One is the gravitational reheating by particle creation in the 
expanding universe in which the beginning of the radiation dominated epoch 
is delayed due to the presence of the epoch of domination of the kinetic 
energy of the inflaton (kination). 
The other is the instant preheating considered by Felder et al. 
in which the Universe becomes radiation dominated soon after the end of 
inflation. We find that the spectrum of the gravitational waves at 
$\sim 100$ MHz is quite sensitive to the reheating process. 
This result is not limited to quintessential inflation 
but applicable to various inflation models.
Conversely, the detection or non-detection of primordial gravitational 
waves at $\sim$100 MHz would provide useful information regarding the 
reheating process in inflation.
\end{abstract}

\maketitle
\newpage
\section{Introduction}
Inflationary paradigm solves various problems in the standard big-bang
model (flatness, horizon, monopole problems, etc.) and predicts
density perturbation and gravitational wave spectra which are both
almost scale-invariant. 
The two of three major predictions of inflation (flat universe, 
scale-invariant density perturbations) are consistent with the current
observations, and the detection of primordial gravitational waves is a
litmus test of the inflationary paradigm. The dynamics of inflation is
believed to be described by a scalar field  called ``inflaton'' (or
multiple of it) with the energy scale of $\sim 10^{15}$GeV.

On the other hand, recent observations indicate that the present
Universe is dominated by dark energy: an (approximately) smooth energy 
component with negative pressure. If dark energy is dynamical, its
dynamics is likely to be described by an ultra-light scalar field with
mass $\siml H_0$, called ``quintessence''. 

Then, it is natural to imagine whether it is possible to 
unify the inflaton and quintessential fields.
A toy model was proposed by Peebles and Vilenkin in which the potential
consists of two parts: $\lambda\phi^4+M^4 ~(\phi \leq 0)$ for inflation
and $\lambda M^8/(\phi^4+M^4)~ (\phi \geq 0)$ for quintessence \cite{pv}. The
two potentials are glued at $\phi=0$ by hand. In this paper, although
not essential, as a slightly elaborated model,
we consider a smooth, exponential type
potential that describes both the inflaton and quintessential
 parts\footnote{This possibility was noted in a footnote of \cite{pv}.}
in the hope that the model could be embedded in particle physics framework.
For other models of quintessential inflation, see \cite{qpote}.

In order to recover the hot big-bang universe, the universe has to be reheated 
after inflation. In models of quintessential inflation proposed so far, 
the only mechanism of reheating that has been considered is
the gravitational particle production in an expanding universe.
The spectrum of gravitational waves based on the gravitational reheating
was calculated \cite{giova1,giova2,bg} and a strong enhancement at $\sim $GHz 
was found (see also \cite{uzan} in different context). 
However, the gravitational reheating process is 
very inefficient and may lead to cosmological problems associated with 
overproduction of dangerous relics such as gravitinos and moduli 
\cite{instant1}. Recently, Felder, Kofman and Linde proposed a much more 
efficient mechanism of reheating
 (called the instant preheating) \cite{instant2}.
In this paper, we calculate the spectra of gravitational waves
in both cases of the instant preheating and gravitational reheating,
and examine the dependence of the gravitational wave spectrum on
the reheating process. 

This paper is organized as follows. 
In Sec. II, we introduce our model of quintessential inflation. 
In Sec. III, we briefly review two reheating processes after quintessential 
inflation. In Sec. IV, we calculate the spectrum of the gravitational 
waves. Sec. V is devoted to conclusion. We use the units of $\hbar =c=1$.

\section{yet another quintessential inflation}

We utilize an exponential potential of the form 
($M_{\rm pl}=1/\sqrt{8\pi G}=2\times 10^{18}$ GeV)
\begin{equation}
V(\phi)=V_0\exp(-\lambda\, \phi/M_{\rm pl}),
\end{equation}
but with $\lambda$ being a function of $\phi$; $\lambda=\lambda(\phi)$.

\paragraph*{Inflation.} 
If the Universe is dominated by a scalar field, then the scale factor
$a(t)$  becomes $a \propto t^{2/\lambda^2}$ and the power-law inflation
is realized for $\lambda < \sqrt{2}$ \cite{power-law}. Recent
observations of cosmic microwave background anisotropies indicate that 
the spectral index of the scalar perturbation, $n_s$, is consistent with 
flat one: $n_s = 0.93 \pm 0.03$ \cite{cmb}. 
Since $n_s-1=-\lambda^2$ under the slow-roll 
approximation, we require 
\begin{equation}
\lambda \lesssim 0.3. \label{condition1}
\end{equation}
If we assume that inflation terminates at
$\phi \simeq 0$ by changing the slope $\lambda$, then the amplitude of
the density perturbations observed by COBE fixes the energy scale of 
inflation: $V_0 \simeq 10^{-12}M_{\rm pl}^4$.

\paragraph*{Kination.} 
After the end of inflation, the universe is still dominated by
$\phi$. Inflation is followed
by a kinetic energy-dominated epoch (hereafter called kination).
During the kination, $\dot\phi^2 \sim V_0(a/a_{kd})^{-6}$ and the scalar 
field evolves as
\begin{equation}
\phi \simeq \sqrt{6}M_{\rm pl}\ln (a/a_{kd}),\label{phi-eq}
\end{equation}
where $a_{kd}$ is the scale factor at the end of inflation. Therefore
in order for the potential energy $V \propto \exp(-\lambda \phi/M_{\rm pl})$ 
not to dominate the energy density of the Universe, it is required that 
\begin{equation}
\lambda > \sqrt{6}. \label{condition2}
\end{equation}

The presence of the epoch of kinetic energy domination is the main
reason of enhancement of amplitude of gravitational waves in
quintessential inflation models discussed below \cite{pv}.

\paragraph*{Dark Energy Domination.} 
The Universe must eventually be dominated again by the scalar field to
account for the acceleration of the Universe. This is accomplished by
the termination of the domination of the kinetic energy of the scalar
field and hence we require 
\begin{equation}
\lambda < \sqrt{6}.\label{condition3}
\end{equation}
The scalar field stops moving when $V(\phi_X) \sim H_0^2 M_{\rm pl}^2$ and 
the Universe becomes eventually dominated by the scalar field again. 

{}From the requirements~(\ref{condition1}), (\ref{condition2}) and 
(\ref{condition3}), we find that different three slopes ($\lambda$) 
are needed for the stages of inflation, kination and dark-energy dominance.
As an example, we consider a potential in 
the following form:
\begin{eqnarray}
&&V(\phi)=V_0\exp(-\lambda (\phi)\phi/M_{\rm pl}),\label{lam}\\
&&\lambda(\phi)=\Bigl(\tanh(\phi/M_{\rm pl})+A\Bigr)\,
{(\phi/M_{\rm pl})^2+u^2\over 
(\phi/M_{\rm pl})^2+v^2}\,,
\label{pot:model}
\end{eqnarray}
where $A,u$ and $v$ are constants. In the following analysis, 
we take $A=1.1$ and arrange $u$ and $v$ to fix $\Omega_{\phi}= 0.7$ today.
Inflation is terminated when $(aH)^{-1}$ is minimum. 
$V_0$ is fixed by the amplitude of density perturbation observed at 
COBE scale : $V^{3/2}/(\sqrt{75}\pi M_{\rm pl}^3 V')=1.9\times 10^{-5}$. 

\paragraph*{Possible Scenario.} 
We speculate on possible scenarios for realizing the above potential
form. The first one is the system of the Lagrangian consisting of 
 a non-canonical and a simple exponential potential:
\begin{equation}
-{1\over 2}Z(\Phi)^2(\nabla \Phi)^2 -V_0\exp(-\alpha \Phi).
\end{equation}
A non-canonical kinetic term appears in supergravity theories
\cite{binetruy} and in higher-dimensional theories via dimensional
reduction. Introducing a new variable $\phi$ such that 
$Z(\Phi)\partial_{\mu}\Phi=\partial_{\mu}\phi$ so that $\phi=\int
Z(\Phi)d\Phi\equiv Y(\Phi)$, we may obtain the
required form of the potential with $\lambda(\phi)\phi=\alpha
Y^{-1}(\phi)$.

Another scenario is a scalar field with non-minimal coupling with
gravity \cite{fm,cy}:
\begin{equation}
f(\phi)R -{1\over 2}(\nabla \phi)^2 - V(\phi).
\end{equation} 
Since the effective potential for the dynamics of the scalar field is
$V(\phi)/f^2(\phi)$ rather than $V(\phi)$ \cite{fm,cy}, we may realize
the required form from a simple form of $V(\phi)$, although we must be
careful about possible time-variation of the gravitational constant and
deviations from general relativity \cite{chiba}.

\section{Reheating processes}

\paragraph*{Gravitational particle production.} 
Massless particles are created due to the change of the expansion-law of the 
space-time. These particles can reheat the universe.
Because the energy of massless particles decreases like that 
of radiation, the energy density of created particles eventually
dominates even if it is initially negligible. 
The created particles thermalize the Universe in the following standard
process \cite{pv}. The energy density of 
the created particles is 
$\rho_m = R H_{kd}^4(a/a_{kd})^{-4}$,
where $H_{kd}$ is the Hubble parameter at the end of inflation and 
$R\sim 10^{-2}$ per scalar field component \cite{ford,damour} 
and $R\simeq 10^{-2}N_s$ for $N_s$ scalar fields. 

The thermalization of $\rho_m$ occurs when the interaction rate
$n\sigma$ is faster than the expansion rate of the Universe $H$. Here 
the number density $n$ and the cross section $\sigma$ are written in terms 
of the energy of created particle 
$\epsilon \sim H_{kd}(a/a_{kd})^{-1}$ as $n\simeq R\epsilon^3$ and
$\sigma \sim \alpha^2\epsilon^{-2}$ with $\alpha\sim 10^{-2}$. Thus the
thermalization occurs at $a_{th}/a_{kd}\sim \alpha^{-1}R^{-1/2}$ and the 
temperature is $T_{th} =R^{3/4} \alpha H_{kd}$.

Since $\rho_m/\rho_{\phi} \sim R H_{kd}^2 M_{pl}^{-2}(a/a_{kd})^{-2}$ during 
 kination, the radiation dominated epoch begins at 
\begin{equation}
a_r/a_{kd}=\sqrt{\frac{3}{R}} H_{kd} M_{\rm pl} \sim 10^6 R^{-1/2} 
\left(\frac{H_{kd}}{10 ^{12} ~\rm {GeV}}\right).
\end{equation}
From this result, we can obtain the temperature at the beginning of the 
radiation.
\begin{equation}
T_r \sim R^{3/4} H_{kd} (3 M _{\rm pl } ^2 )^{-1/2} 
\sim 10^6 R^{3/4} \left(\frac{H_{kd}}{10 ^{12} ~\rm {GeV}}\right) ~{\rm GeV}.
\end{equation}
This temperature is sufficient for the beginning of the standard hot 
big bang model. A typical result of the time evolution is given in Fig. 1.

\paragraph*{Instant preheating.}
We consider the so-called instant preheating as another reheating mechanism.
Felder et al. proposed a much more efficient mechanism of reheating which 
works even for a potential without a minimum \cite{instant1}. 
The mechanism is similar to preheating by parametric resonance \cite{instant2}.
They introduced the interaction Lagrangian,
$-\frac{1}{2}g^2 \phi ^2 \chi ^2 -\frac{1}{2} \psi \bar \psi \chi$, 
where $g$ and $h$ are the coupling constants. 
After the end of inflation, the energy of the inflaton $\phi $ is 
transferred to the scalar field $\chi $ and the field $\chi $ decays to 
the fermions $\psi $. 

For simplicity, we assume that inflation ends at $\phi =0$ and the scalar 
field $\chi $ does not have a bare mass.
Then, the effective mass of the field $\chi $ is $m_\chi =g|\phi |$. 
The particle creation of the field $\chi $ occurs when the adiabaticity 
condition is violated, $|\dot m_\chi|\simg m _\chi ^2$.
This happens right after the kination begins.
Hence when
\beq
|\phi|\siml \phi_*=\sqrt{|\dot\phi_{kd}|/g}\simeq V_0^{1/4}/\sqrt{g},
\eeq
where $\dot\phi_{kd}$ is $\dot\phi$ at the end of inflation. 
In order for the particle creation to be completed
within a sufficiently short period, we must have
$\phi <M_{\rm pl}$ and $g$ must satisfy $g > 10^{-6}$.
The time interval of the particle production is then written as
\beq
\Delta t \sim \frac{\phi _*}{| \dot \phi _{\rm kd} |} 
\simeq g^{-1/2}V_0^{-1/4}.
\eeq
Thus the created particles will have typical momenta
$k \sim {\Delta t}^{-1} \simeq g^{1/2}V_0^{1/4}$ \cite{instant1}.
The occupation number of the created particles is 
$n(k) = \exp \left(-{\pi k^2/g \sqrt{V _0}}\right)$ \cite{instant2}.
Integrating this equation, we can get the number density of the field $\chi $
\begin{equation}
n _\chi =\frac{1}{2 \pi ^2}\int ^\infty _0 dk k^2 n (k) =
\frac{\left(g \sqrt{V _0}\right)^{3/2}}{8 \pi ^3}.
\label{suumitudo}
\end{equation}
The energy density immediately after the creation is 
$\rho _{\chi} \sim {\left(g \sqrt{V _0}\right)^{2}}/{8 \pi ^3}$.
Soon after that, the field $\chi $ decays into the fermion $\psi $ with 
decay rate $\Gamma = h^2 m _\chi/8 \pi=h^2 g |\phi | /8 \pi$. 
The subsequent decay of the field $\chi $ leads to a complete reheating of the Universe.

The equation of the field $\phi$ after particle production is written as
\begin{equation}
\ddot \phi + 3 H \dot \phi + g^2 \langle \chi ^2 \rangle \phi=0,
\end{equation}
where the potential can be neglected in the kination.
When the field $\phi$ becomes greater than $\phi_*$ and particle 
production ends, the particles $\chi$ become nonrelativistic. 
Then we can calculate $\langle \chi ^2 \rangle$
\begin{equation}
\langle \chi ^2 \rangle \approx \frac{1}{2 \pi ^2} \int \frac{n _k k^2 dk}
{\sqrt{k^2 + g^2 \phi ^2}}
\approx \frac{n _\chi }{g \phi } \approx \frac{(g \sqrt{V_0} )^{3/2}}
{8 \pi ^3 g \phi }\left({a_{kd}\over a}\right)^3.
\end{equation}
Therefore the backreaction of created particles on $\phi$ is negligible 
if $g^2 \left\langle \chi^2 \right\rangle \phi <3 H \dot\phi $, that is  
\begin{equation}
a<a_{kd}\left(\frac{8 \sqrt{6} \pi ^3 V _0 ^{1/4} }
{g ^{5/2} M_{\rm pl}}\right)^{1/3}.
\label{joukenne}
\end{equation}
We consider the case of $a_b/a_{kd}>1$, so we get $g<0.8$ from 
Eq.(\ref{joukenne}).
If $\Gamma >H_b$, where $H_b$ is the Hubble parameter 
at the scale factor $a_b = a_{kd}[8 \sqrt{6} \pi ^3 V _0 ^{1/4} /
(g ^{5/2} M_{\rm pl})]^{1/3}$,
particles $\chi $ will decay to fermions $\psi $ at $a<a_b$ 
and the force driving the field $\phi $ back to $\phi =0$ will disappear.
This condition is given by
\begin{equation}
h^2 > \frac{g^{3/2} V_0 ^{1/4}}{3 \sqrt{2} \pi ^2 M_{\rm pl}},
\label{hgcondition}
\end{equation}
where we have assumed $\phi \sim M_{\rm pl}$. 
When the backreaction is nonnegligible, the scalar field $\chi $ dominated 
epoch begins. 
The reason is that the condition~(\ref{joukenne}) is essentially
the same as that for $\rho _\chi <\rho _\phi $.
When $H \sim \Gamma $, the field $\chi $ decays into $\psi$ 
and the energy is transferred to the radiation energy. Thus 
the radiation dominated epoch starts. In Fig. 2, we show the results of 
time evolution of fields.

\begin{figure}[tbp]
  \begin{center}
    \includegraphics[keepaspectratio=true,height=50mm]{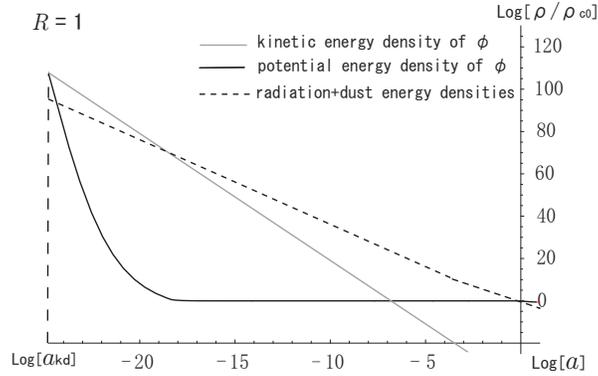}
  \end{center}
  \caption{Evolution of the energy densities for gravitational reheating.
  The solid line is the potential energy density of $\phi$. The gray 
line is the kinetic energy density of $\phi $.
  The dotted line is the radiation and dust energy densities.}
  \label{R=1ene.eps}
\end{figure}

\begin{figure}[tbp]
  \leavevmode
  \begin{center}
    \begin{tabular}{ c c }
 \includegraphics[keepaspectratio=true,height=50mm]{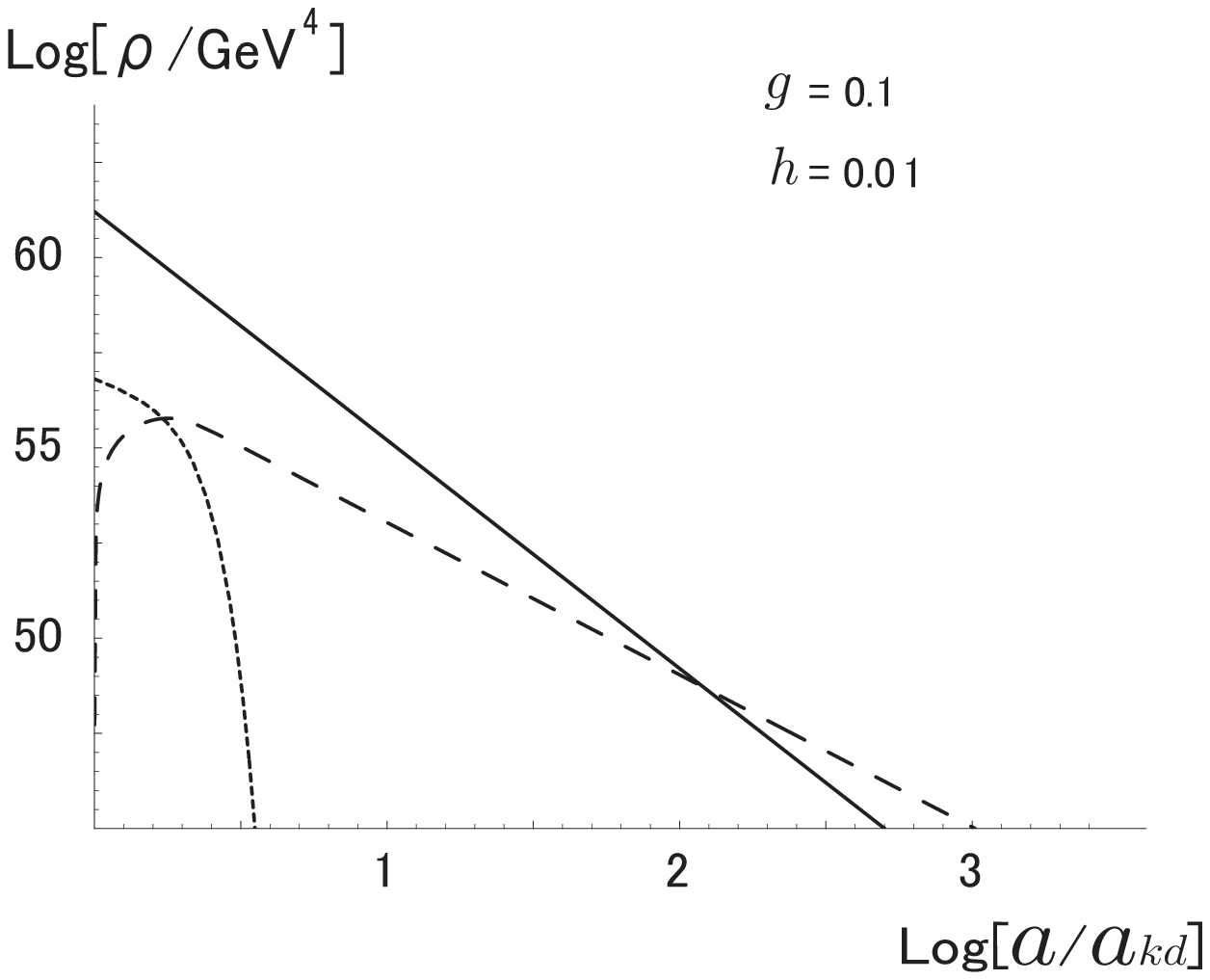}
      &
 \includegraphics[keepaspectratio=true,height=50mm]{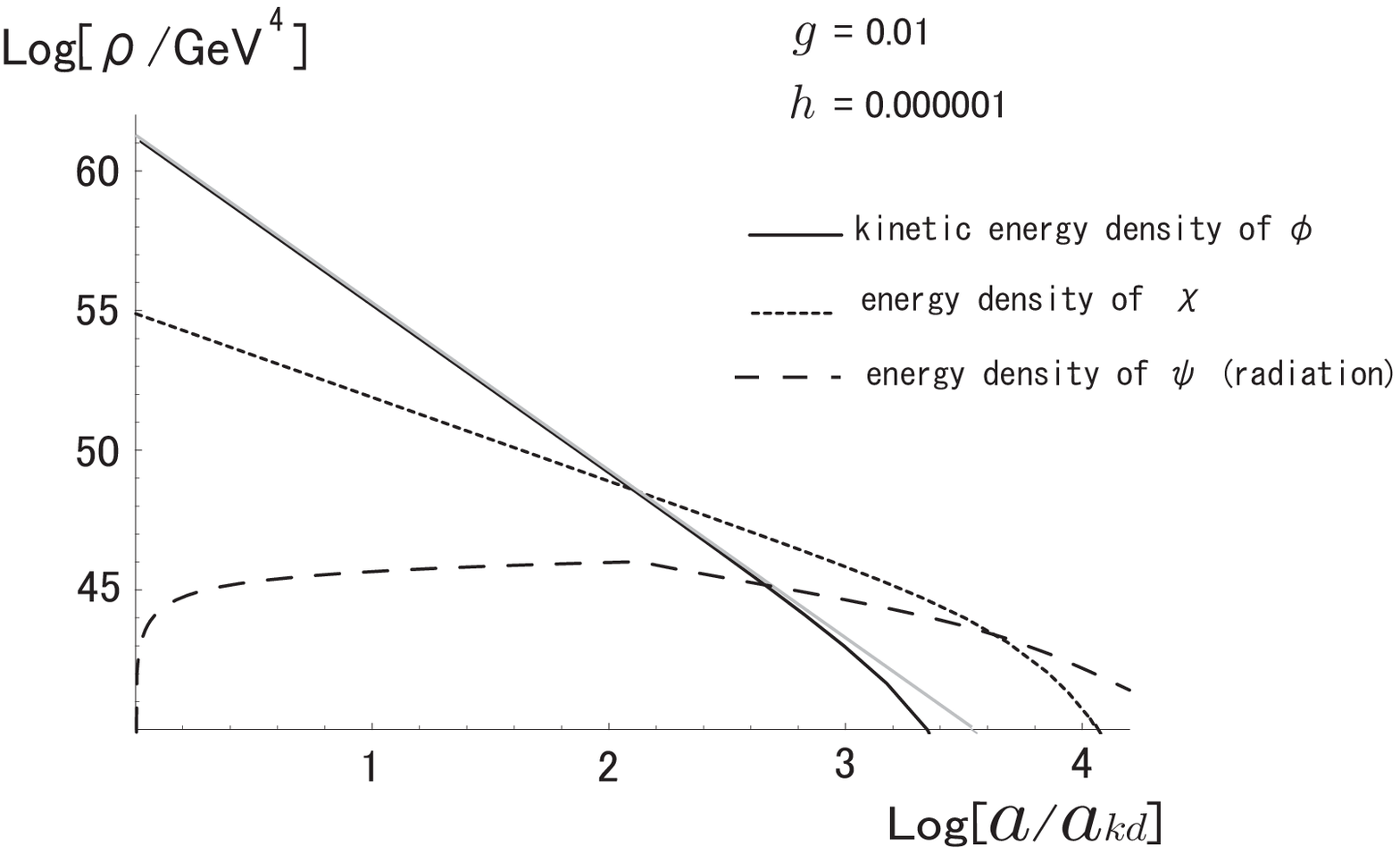}
          \end{tabular}
 \caption{Evolution of the energy densities for instant preheating.
          The solid line is the kinetic energy density of $\phi $, the dotted 
          line is the energy density 
  and the dashed line is energy density of $\psi $.
 In the right panel there is the era dominated by the scalar field $\chi $,
 because the decay rate of the field $\chi $ is low.}
   \label{reheating}
   \end{center}
\end{figure}

\section{Gravitational waves}
In each reheating process, we take $A=1.1$ in Eq.(\ref{pot:model}) and 
arrange $u,~v$ to fix $\Omega_{\phi}= 0.7$ at the present.
From this result, we obtain the scale factor and the Hubble parameter 
at the beginning of each epoch.

We calculate the graviton spectrum produced by an inflationary stage by
using the Bogoliubov transformation
in terms of conformal time $\tau$ \cite{allen,curve}.
But the spectrum shape is easily understood in the following way before 
carrying out a detailed calculation.
When the gravitational waves come out of the horizon during inflation, 
the amplitude of the gravitational waves is
frozen and the value is proportional to $H_{out}$, where $H_{out}$ is 
the Hubble parameter when the scale of the perturbation crosses
the Hubble horizon.
After reentering the horizon at scale factor $a_{in}$, the amplitude 
decreases as $1/a$.
So the today characteristic amplitude $h_c (\omega )$ 
which reentered the horizon in an era with scale factor $a \propto t ^n$ is
\beq
h _c \propto {H _{out}}\frac{a _{in}}{a _0}
\propto f^{(3/2)-\nu + (n/n-1)}.  
\eeq
where 
\begin{eqnarray}
\nu =\frac{1}{2}\left(1+ \frac{4}{2-\lambda ^2}\right)\,,
\end{eqnarray}
and $\lambda $ here is its value during inflation
which we approximate by $\lambda (\phi \rightarrow -\infty)$ in Eq.~(\ref{lam}) 
for simplicity. This means that the scale factor during inflation is 
approximated to be proportional to $t^{2/\lambda^2}$. 
{}From this, the energy density of the gravitational waves
per unit logarithmic interval of frequency is
\begin{equation}
 \Omega _{\rm gw}(f) = \frac{1}{\rho _c}\frac{d\rho _{\rm gw}}{d \log f}=
\frac{1}{6 H _0 ^2}f ^2
h _c ^2 (f ) \propto f ^{3-2\nu + (2n/n-1)+2},
\end{equation}
where $\rho_c$ is the critical energy density.
Our model consists of the five phases: inflationary phase (IN), 
kinetic-energy dominated phase (KD), radiation dominated phase (RD), 
matter dominated phase (MD) and
potential dominated phase (PD). Among them, 
the phases when the gravitational waves 
reenter the horizon are 
KD ($n=1/3$), RD ($n=1/2$) and MD ($n=2/3$).
Correspondingly, the spectrum has three branches which are proportional
to $f^{1+(3-2\nu)}$, $f^{3-2 \nu}$ and $f^{-2+(3-2 \nu)}$ 
and the shapes are showed in Fig. 3.

The detailed calculation using the Bogoliubov transformation is 
as follows.
The scale factor in each phase is given by
\begin{eqnarray}
&&a(\tau ) = a _{kd} \left(\frac{\tau }{\tau _{kd} }\right)^{{(1-2 \nu )}/2}
\hspace{8.4cm}\mathrm{for}~
\mathrm{IN},
\nonumber\\[3mm]
&&a(\tau ) = a _{kd}
\left[
(1-2 \nu )
\left(
\frac{\tau }{\tau _{kd}}+\frac{2 \nu }{1-2 \nu }\right)
\right]^{1/2} 
\hspace{5.5cm}\mathrm{for}~\mathrm{KD},
\nonumber\\[3mm]
&&a(\tau ) =\frac{a _{kd} ^2}{a_r}\frac{1-2 \nu }{2}\frac{\tau
_{r}}{\tau _{kd}}
\left(
\frac{\tau }{\tau _{r}} +\frac{a _{r} ^2}{a _{kd} ^2}\frac{2}{1-2 \nu }
\frac{\tau _{kd}}{\tau _{r}}-1
\right)
\hspace{4cm}~\mathrm{for}~\mathrm{RD},
\nonumber\\[3mm]
&&a(\tau ) = a_d
\left(\frac{1-2 \nu }{4}\frac{a _{kd} ^2}{ a _d a_r}\frac{\tau _d}{\tau
_{kd}}
\right)^2
\left[
\frac{\tau }{\tau _d} +\frac{4}{1-2 \nu }\frac{a _d a_r}{a _{kd} ^2}\frac{\tau
_{kd}}{\tau _d}-1
\right]^2
\hspace{2.5cm}\mathrm{for}~\mathrm{MD},
\nonumber\\[3mm]
&&a(\tau ) =\frac{a_p}{2}
\frac{ \sqrt{a _p}}{\sqrt{a _p }- \sqrt{a _d}}
\left(
1-\frac{\tau _d}{\tau _p}
\right)
\left[
\frac{1}{2}
\frac{\sqrt{a _p}}{\sqrt{a _p} - \sqrt{a _d}}
\left(
1-\frac{\tau _d}{\tau _p}
\right)
+1-\frac{\tau }{\tau _p}
\right]^{-1}
\quad\mathrm{for}~\mathrm{PD},
\end{eqnarray}
where subscripts ${kd}$,~$r$,~$d$ and $p$ mean the epochs of
the phase transitions between IN-KD, KD-RD, RD-MD and MD-PD stages,
respectively.
In the PD phase, we have assumed that the universe accelerates exponentially,
since the potential energy density is nearly constant as seen from
Fig.~\ref{R=1ene.eps}.
The conformal time at each phase transition is expressed as
\begin{eqnarray}
&&\tau _{kd} =
\frac{1-2 \nu}{2}\frac{1}{a _{kd} H _{kd}}
,\nonumber\\
&&\tau _{r} =
\left(
\frac{a _{r} ^2}{a _{kd} ^2}-1
\right)
\frac{1}{2 a _{kd} H _{kd}}+\tau _{kd}
,\nonumber\\
&&\tau _d =
\left(
\frac{a _d}{a _{r}}-1
\right)
\frac{1}{a _{r} H _{r}} + \tau _{r}
,\nonumber\\
&&\tau _{p} =
\left(
\sqrt{\frac{a _{p}}{a _{d}}}-1
\right)
\frac{2}{a _{d} H _{d}}+\tau _{d}.
\label{tauh}
\end{eqnarray}
Note that $\tau_{kd}<0$.

In the TT (transverse-traceless) gauge, 
a positive frequency function for the
gravitational perturbation with comoving 
wave number {$\bf k$} is represented by
\begin{equation}
{h_{ij} ^{(\sigma,{\bf k})}}
= {a(\tau )}^2 {e^{(\sigma)}_{ij}}({\bf k}) {{\mu  }_k}(\tau ) 
e^{i {\bf k \cdot x}},
\end{equation}
where $e^{(\sigma)}_{ij}({\bf k})$ ($\sigma=1,2$)
are two independent polarization tensors normalized as
\begin{eqnarray}
e^{(\sigma)\,ij}({\bf k})e^{(\sigma')}_{ij}({\bf k})
=\delta^{\sigma\sigma'}\,.
\end{eqnarray}
The Einstein equations give an equation for ${{\mu}_k}$
\begin{equation}
{{\mu }_k}'' + 2 \frac{a'}{a}{\mu _k}' +k^2{{\mu }_k}=0,
\label{gravitational-eq}
\end{equation}
where the prime denotes the derivative with respect to the conformal time.
The mode function $\mu$ satisfies the normalization
\begin{equation}
\mu^*  \mu ' -\mu {\mu^*}' =- i /a^2. 
\label{normalization}
\end{equation}
In the inflationary phase, assuming the standard Bunch-Davies vacuum,
the solution is given by
\begin{eqnarray}
&&\mu _{i} (\tau ) =
\frac{\sqrt{\pi }}{2 {a _{kd}}}
(-\tau _{kd})^{\frac{1}{2}-\nu} (-\tau)^\nu
H _{\nu } ^{(1)}( -k \tau ),
\end{eqnarray}
where the subscript $i$ means it is the solution
in the inflation phase and $H _{\nu } ^{(1)}(x)$ is the 
Hankel function of the first kind.

The solution in the kination is 
\begin{eqnarray}
&&\mu _{kd} (\tau ) = \alpha _{kd} \psi _{kd}(\tau)+ \beta \psi^* _{kd}(\tau)
\\
&&\psi _{kd}(\tau) =\frac{\sqrt{\pi }}{2 {a _{kd}}} 
\left(\frac{1}{1-2\nu} \tau _{kd} \right)^{1/2}
H _0 ^{(2)} \left( k \left[\tau +\frac{2 \nu }{1-2\nu }\tau _{kd}\right] 
\right),
\end{eqnarray}
where $\alpha _{kd}$ and $\beta _{kd}$ are called the Bogoliubov 
coefficients, $\psi _{kd} $ is the mode function in the kination 
normalized as Eq. (\ref{normalization}),
and $H_0^{(2)}(x)$ is the Hankel function of the second kind.
The Bogoliubov coefficients are determined by the condition that
${{\mu }}$ and ${{\mu}}'$ be continuous at the phase transition.
From this condition at $\tau =\tau_{kd}$ we get
\begin{eqnarray}
&&\alpha _{kd} = \frac{1}{H _\nu  ^{(1)} (-k \tau _{kd})}
\left[
\left( \frac{1}{1-2\nu}
\right)^{1/2}
H _\nu ^{(1)} (-k \tau _{kd})
- \beta _k H _0 ^{(1)} \left( \frac{1}{1-2\nu}k \tau _{kd} \right)
\right],
\nonumber
\\
&&\beta _{kd} = \frac{i \pi}{8} \left( \frac{4}{1-2\nu}\right)^{1/2}
\tau _{kd}
\nonumber\\
 & &
~~\qquad\times \left[
-\frac{k}{2}H_1 ^{(2)}\left(\frac{1}{1-2\nu} k \tau _{kd}\right)
 H _\nu  ^{(1)} (-k \tau _{kd})- 
\nu \tau _{kd} ^{-1} H _\nu  ^{(1)} (-k \tau _{kd}) H _0 ^{(2)}
\left(\frac{1}{1-2\nu} k \tau _{kd}\right)\right.
\nonumber\\
 & & 
~~\qquad\left. + \frac{k}{2}H _0 ^{(2)}\left(\frac{1}{1-2\nu}k 
\tau _{kd}\right) \left(H_{\nu -1} ^{(1)} (-k \tau _{kd}) - 
H_{\nu +1} ^{(1)} (-k \tau _{kd})\right)\right].
\label{bogo-kination}
\end{eqnarray}
For wavelengths of cosmological interest, we have
$-k \tau_{kd}\ll 1$. When we assume $-k \tau_{kd} \ll 1$, we get
\begin{equation}
\beta _{kd}
=\frac{\Gamma  (\nu)}{\pi}\left(\frac{1-2\nu}{4} \right)^{1/2}
\left[\left(\frac{4}{1-2\nu} \right)\frac{a_{kd} H_{kd}}{k}\right]^\nu,
\end{equation}
where we have replaced $\tau_{kd}$ by its expression in terms
of the scale factor and the Hubble parameter given in Eq.(\ref{tauh}).

Here we need to mention the adiabatic theorem \cite{allen,curve}.
The transition of the phases is not instantaneous in reality.
The change of the cosmological evolution occurred with 
the time scale $\Delta T=H^{-1}$.
The time scale associated with the mode function with the
comoving wavenumber $\bf k$ is the frequency 
$\omega = k /a$.
If $k /a \ge \Delta T=H^{-1}$, this change of phase is adiabatic and no 
particle creation occurs.
This results in $\alpha =1 $, $\beta =0$ for $k \ge a H$
at an epoch of a phase transition.

In order to obtain the spectrum of the present gravitational waves
we need the solution in the potential dominated epoch.
It is acquired by the the Bogoliubov transformation.
The relation between the mode functions are written
using the Bogoliubov coefficients. 
\begin{eqnarray}
&&\psi_{kd}  = \alpha_r \psi _{r}  + \beta_r \psi ^* _r,
\nonumber\\
&&\psi_{r}  = \alpha_d \psi _{d}  + \beta_d \psi ^* _d,
\nonumber\\
&&\psi_{d}  = \alpha_p \psi _{p}  + \beta_p \psi ^* _p.
\end{eqnarray}
These Bogoliubov coefficients are obtained by requiring the continuity 
condition at each phase transition.
The mode function at the PD epoch is expressed as
\begin{equation}
\mu_p =\alpha_{total }~ \psi_{p}+ \beta_{total }~ \psi ^* _p,
\end{equation}
where $\alpha_{total }$ and $\beta_{total }$ are given by
\begin{eqnarray}
&&\left(\begin{array}{rr}
\alpha _{total}& \beta _{total}\\
\beta ^* _{total}& \alpha ^* _{total}
\end{array}\right)
=\left(\begin{array}{rr}
\alpha _{kd}& \beta _{kd}\\
\beta ^* _{kd}& \alpha ^* _{kd}
\end{array}\right)
\left(\begin{array}{rr}
\alpha _r& \beta _r\\
\beta ^* _r& \alpha ^* _r
\end{array}\right)
\left(\begin{array}{rr}
\alpha _d& \beta _d\\
\beta ^* _d& \alpha ^* _d
\end{array}\right)
\left(\begin{array}{rr}
\alpha _{p}& \beta _{p}\\
\beta ^* _{p}& \alpha ^* _{p}
\end{array}\right).
\nonumber\\
& & {}
\end{eqnarray}
In the PD epoch, the wavelength cannot come into the horizon because of the 
accelerated expansion.  So, for all scales which we can observe 
($k>a_0 H_0$), we have $\alpha _p =1$ and $\beta_p =0$.

The number density of created gravitons with conformal wave number $\bf k $ is
\begin{equation}
N(k)= |\beta _{total}| ^2 .
\end{equation}
The energy density of gravitational waves with frequency 
$f= k /2 \pi a_0$ is written taking into account of two polarization 
of the gravitational waves as
\begin{equation}
d\rho_{\rm gw} (f ) = 2 ~(2\pi  f) ~|\beta _{\rm total}|^2    4 \pi f ^2 df. 
\end{equation}
{}From this the density parameter of gravitational waves is
\begin{equation}
 \Omega _{\rm gw}(f) = 16 \pi^2 \rho _c ^{-1}|\beta _{\rm total}|^2 f^4 .
\end{equation}
Thus we finally obtain 
\begin{eqnarray}
&&\Omega _{\rm gw}(f) =\frac{4(1-2 \nu) \Gamma^2 (\nu)}{3 M_{\rm pl} ^2 
H _0 ^2} \left[\frac{2 H _{kd}}{\pi (1- 2\nu)}
\frac{a _{kd} }{a_0} \right]^{2 \nu} f^{1 +(3-2 \nu)}  ,~~~~~~~~
\frac{H_r}{2 \pi }\frac{a_r}{a_0}<f<\frac{H_{kd}}{2 \pi }\frac{a_{kd}}{a_0},
\\
&&\Omega _{\rm gw}(f) =\frac{2(1-2 \nu) \Gamma^2 (\nu)}{3\pi M_{\rm pl} ^2 
H _0 ^2} \left[\frac{2 H _{kd}}{\pi (1- 2\nu)}
\frac{a _{kd} }{a_0} \right]^{2 \nu}  H_r \frac{a_r}{a_0}f^{(3-2 \nu)},~~~
\frac{H_d}{2 \pi }\frac{a_d}{a_0}<f<\frac{H_{r}}{2 \pi }\frac{a_r}{a_0},
\\
&&\Omega _{\rm gw}(f) =\frac{(1-2 \nu) \Gamma^2 (\nu)}{6 \pi^3 M_{\rm pl} ^2 
H _0 ^2} \left[\frac{2 H _{kd}}{\pi (1- 2\nu)}
\frac{a _{kd} }{a_0} \right]^{2 \nu} \left(H _d \frac{a _d }{a_0} \right)^2 
H_r \frac{a_r}{a_0}  f^{-2 +(3-2 \nu)}, 
\nonumber\\
&&~~~~~~~~~~~~~~~~~~~~~~~~~~~~~~~~~~~~~~~~~~~~~~~~~~~~~~~~~~~~~~~~~~~~~~~~~~~~
\frac{H_0}{2 \pi }<f<\frac{H_{d}}{2 \pi }\frac{a_{d}}{a_0}.
\end{eqnarray}
Note that $\nu =3/2$ corresponds to de Sitter inflation.

\begin{figure}[tbp]
  \leavevmode
  \begin{center}
    \begin{tabular}{ c c }
 \includegraphics[keepaspectratio=true,height=50mm]{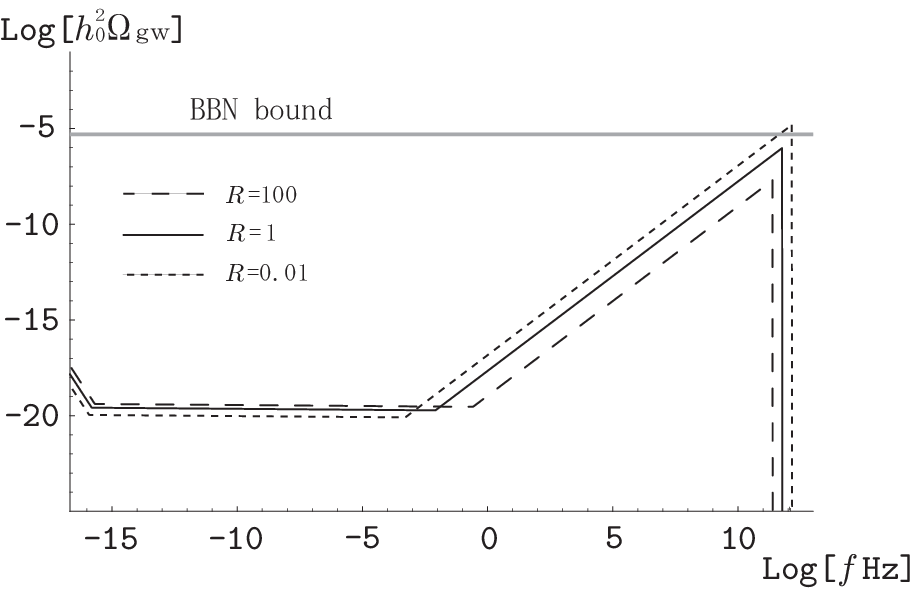}
      &
 \includegraphics[keepaspectratio=true,height=50mm]{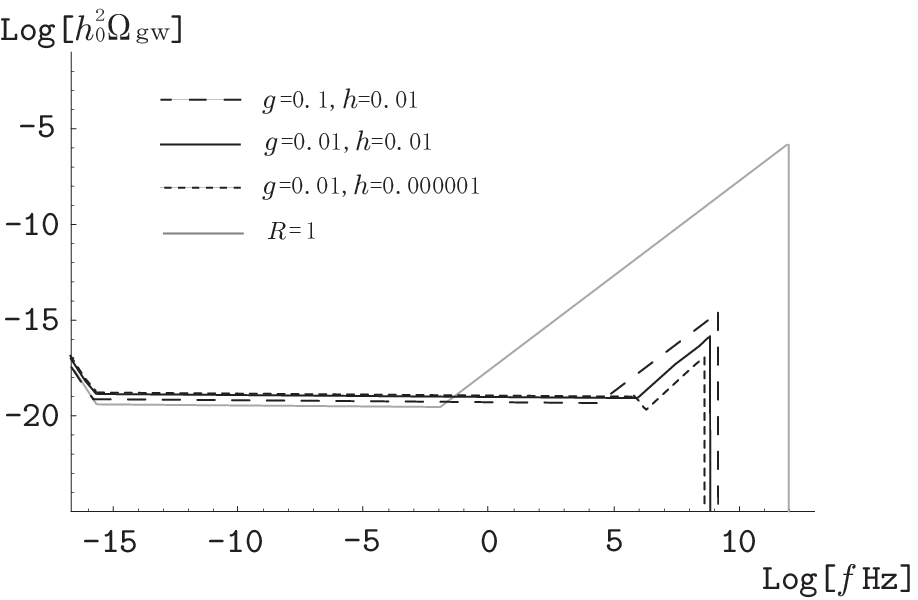}
          \end{tabular}
 \caption{The spectrum of gravitational waves. 
 The left panel is in the case of the reheating by gravitational particle 
production.
The right one is in the case of instant preheating. For comparison, 
we also plot the result of gravitational reheating with $R=1$.}
 \label{fig:spectrum.eps}
  \end{center}
\end{figure}

The spectrum of gravitational waves for each reheating process is shown 
in Fig. 3. The effect of kination shows up in the band
$a _r H _r / 2 \pi a_0 < f< a _{kd} H _{kd} / 2 \pi a_0$.
The spectrum in this band is not flat but proportional to $f^{1 +(3-2 \nu)}$.
Therefore the gravitational waves at high frequency are amplified much 
more than that in ordinary inflation models.
The duration of kination depends on the efficiency of reheating.
The longer the duration is, the larger the amplitude becomes. 
The gravitational particle production is a least efficient reheating
mechanism, which may give rise to the amplitude of gravitational waves
large enough to be detectable. 
However the energy density of the gravitational waves is constrained by 
the success of the Big Bang nucleosynthesis, 
$\Omega_{\rm gw} < \Omega_{\rm gw}^{BBN}= 5 \times 10^{-6}h _0 ^{-2}$ 
\cite{maggy}. Hence the efficiency $R$ of the reheating by gravitational 
particle production has a lower limit,
\beq 
R \simg 0.2 (\Omega_{\rm gw}^{BBN}/5 \times 10^{-6}h _0 ^{-2})^{-1} 
(H _{kd}/10 ^{12} ~\rm {GeV})^4.
\eeq 
The sensitivity of the advanced LIGO will be 
$\Omega _{\rm gw} \simeq 10^{-11}$ at $10 \sim 100$Hz and 
that of LISA will $\Omega _{\rm gw} \simeq 10^{-11}$ at 
$1$mHz \cite{maggy}. 
Thus the detectability gravitational waves by these experiments 
appears unlikely. 

When we calculate the gravitational wave spectrum 
we assumed that $\nu $ is constant in the inflationary phase.
This is because $\nu $ changes rapidly at the end of inflation in our toy 
model. However, if we consider a model that $\nu $ changes mildly, 
we must consider corresponding effects on the spectrum.
If $\nu $ is larger than 3/2 in the smaller scales than the COBE bound region 
($\Omega _{\rm gw}<7 \times 10^{-11}(H_0/f)^2 h_0 ^{-2},~10^{-18}{\rm Hz}<f<10^{-16}{\rm Hz}$ \cite{maggy}),
the slope in the bands of the kination becomes looser and 
the energy density of the gravitational waves may be high enough 
in the detectable bands by LIGO and LISA. 

Instant preheating is so efficient that the duration of kination is much 
shorter than that in gravitational reheating. This results in suppression of 
the enhancement of the spectrum of the gravitational waves at $\sim$ 100MHz. 
Hence we find that the spectrum of the gravitational waves is quite sensitive
to the reheating process after inflation. 
From Fig. 3 we may obtain information regarding the thermal history of the 
universe after inflation from the spectrum of the gravitational waves at 
$\sim $100MHz.

\section{Conclusion}
We investigated the possibility of quintessential inflation by
an exponential-type potential. In this inflation model,
the present accelerated expansion is explained by the same scalar field 
that caused inflation in the past.
We considered two extremes of reheating process:
the gravitational particle production which is least efficient
and the instant preheating which is most efficient and has not been considered 
in the context of quintessential inflation. 
We calculated the spectrum of gravitational waves 
produced during inflation.
We found that the spectrum of the gravitational waves is quite sensitive
to the reheating process. Less efficient reheating results 
in larger amplitude of the spectrum. 
We emphasize that these results are not limited to quintessential inflation and 
insensitive to the detailed model of inflation since the spectrum is
essentially determined by the expansion rate during
inflation and that at the horizon-crossing time. 
On very large scales corresponding to the Hubble rate 
at an early stage of inflation,
it is severely constrained by CMB observations, while on very small scales,
the dominant constraint comes from BBN which is less stringent.
Therefore, 
while it is definitely important to obtain the shape of the spectrum at 
high frequencies ($\sim$ 100 Hz) or at very high frequencies ($\sim$100 MHz), 
even the detection or nondetection of the gravitational waves
would provide us with 
useful information of the early universe.

In particular, for detectors with arm length of 1 m, for example,
the typical frequency 
range would be 100 MHz. Such tabletop gravitational wave detectors
may be constructed with much less cost \cite{kawamura}. 
The implications of such high frequency detectors on the early universe
would be enormous. 

\acknowledgments
The authors would like to thank the Yukawa Institute for Theoretical 
Physics at Kyoto University, where this work was initiated during 
the workshop on gravitational waves (YITP-W-01-16). 
This work was supported in part by a Grant-in-Aid for Scientific 
Research (No.~15740152) from the Japan Society for the Promotion of
Science and by a Grant-in-Aid for Scientific Research on Priority Areas 
(Nos.~14047212 and 14047214) from the Ministry of Education, Science,
 Sports and Culture, Japan.

\end{document}